\documentclass[conference]{IEEEtran}
\usepackage{graphicx}
\usepackage{url}
\usepackage{amsmath}
\usepackage{amssymb}
\usepackage{tcolorbox}
\usepackage[utf8]{inputenc}
\usepackage[T1]{fontenc}
\usepackage[T2A]{fontenc}
\usepackage[russian,english]{babel}
\usepackage{enumitem}
\begin{document}

\title{Microservice Architecture Patterns for Scalable Machine Learning Systems}
\author{Sowjanya Karanam, Jayanth Bhargav}
\maketitle

\begin{abstract}
Machine learning is now a central part of how modern systems are built and used, powering everything from personalized recommendations to large-scale business analytics. As its role grows, organizations are facing new challenges in managing, deploying, and scaling these models efficiently. One approach that has gained wide adoption is the use of microservice architectures, which break complex machine learning systems into smaller, independent parts that can be built, updated, and scaled on their own. In this paper, we review how major companies such as Netflix, Uber, and Google use microservices to handle key machine learning tasks like training, deployment, and monitoring. We discuss the main challenges involved in designing such systems and explore how microservices fit into large-scale applications, particularly in recommendation systems. We also present some simulation studies showing that microservice-based designs can reduce latency and improve scalability, leading to faster, more efficient, and more responsive machine learning applications in real-world and large-scale systems.

\end{abstract}

\section{Introduction}
As machine learning becomes part of more products and services, companies need ways to make their systems easier to grow, maintain, and improve~\cite{oye2024microservices}. When everything is built as one large system, even small changes can be slow and difficult. Microservices help by breaking machine learning workflows into smaller, independent parts, such as data handling, model training, serving, and monitoring. Each part can be updated or scaled without affecting the rest of the system. This makes development faster and more reliable. In this paper, we look at how microservice design patterns are used in real companies to build and manage large-scale machine learning systems more efficiently. The loosely coupled nature of microservices offers several advantages. Each service can be designed and deployed independently, giving development teams the freedom to make changes quickly without affecting the rest of the system~\cite{palamarchuk2022methods}. Because services communicate through well-defined interfaces, they can be built using different programming languages, frameworks, or environments, allowing developers to choose the most suitable tools for each task~\cite{shabani2022design}. Microservices also work naturally with container-based technologies like Docker and orchestration tools such as Kubernetes or Mesos, making it easier to scale, automate, and manage deployments~\cite{de2019monolithic}. By keeping each service focused on a single, clear responsibility, the codebase becomes more organized, easier to understand, and simpler to maintain~\cite{blinowski2022monolithic}. Another key benefit is fault isolation, i.e., if one service fails, the rest of the system continues to operate normally, improving reliability and making the overall architecture more resilient~\cite{blinowski2022monolithic}.

\section{Related Work}

In this section, we look at several studies that explore how microservice architectures are being used to improve the design and performance of machine learning systems. Earlier work points out that traditional, monolithic ML setups often struggle as data grows and models become more complex, leading to delays in training, deployment, and system maintenance. To address these challenges, many researchers and companies have started breaking machine learning workflows into smaller, specialized services that handle specific tasks such as data preparation, feature extraction, model training, and prediction. The studies we review highlight how this approach, supported by containerization and orchestration tools, has helped build faster, more reliable, and easier-to-manage ML systems in real-world environments. Many organizations are adopting microservice-based architectures for their machine learning systems~\cite{oye2024microservices}. In this approach, each stage of the ML lifecycle, like data preparation, model training, serving, and monitoring, is developed and deployed as an independent service. This separation allows teams to build, test, and improve each component without disrupting other parts of the workflow. It also supports parallel development, modular scalability, and easier fault isolation, which are essential for maintaining reliability in large-scale production environments.

Major technology companies such as Google, Uber, and Netflix have successfully adopted this approach to scale their machine learning operations. Google’s TensorFlow Extended (TFX) platform decomposes the entire machine learning workflow into integrated microservices responsible for data ingestion, validation, transformation, model training, evaluation, and serving~\cite{baylor2017tfx}. This modular setup allows Google to maintain high reliability while enabling rapid experimentation and deployment. Similarly, Uber’s Michelangelo platform divides its machine learning functionality into services for feature generation, model training, serving, and monitoring~\cite{hermann2018scaling}. This enables Uber to support both online and offline pipelines efficiently and perform large-scale model retraining without interfering with real-time operations.

\begin{table*}[!t]
\centering
\caption{Literature Review on Microservices in Large-Scale Machine Learning Applications}
\renewcommand{\arraystretch}{1.4}
\begin{tabular}{|p{4cm}|p{8cm}|p{1.5cm}|}
\hline
\textbf{Primary Application} & \textbf{Uses / Role of Microservice} & \textbf{References} \\
\hline
Google TFX (TensorFlow Extended) & Modular ML pipeline where each stage, i.e., data ingestion, validation, transformation, training, and serving is deployed as an independent microservice orchestrated using Apache Beam and Kubernetes. Ensures reproducibility, scalability, and automation of end-to-end ML workflows. & \cite{baylor2017tfx} \\
\hline
Uber Michelangelo Platform & Divides the ML lifecycle into services for feature generation, model training, deployment, and monitoring. Supports online/offline pipelines and provides APIs for model serving and A/B testing. & \cite{olowoniyianalyzing} \\
\hline
Netflix Recommendation System & Uses microservices to separate data pipelines, nearline event computation, and real-time model inference. Each service scales independently for recommendation generation, personalization, and monitoring. & \cite{amatriain2015recommender} \\
\hline
Amazon SageMaker & Provides container-based microservices for training, inference, and model hosting. Each component runs as an isolated service, allowing rapid deployment and scaling in the AWS ecosystem. & \cite{singh2024development} \\
\hline
Microsoft Azure ML Service & Implements microservices for data preparation, model management, and deployment through Docker and Kubernetes. Enables CI/CD for ML workflows and supports cross-language integration. & \cite{barnes2015microsoft}\\
\hline
Kubeflow Pipelines & Open-source framework that uses microservices to define, schedule, and monitor ML workflows on Kubernetes. Each pipeline component runs as a containerized service for reproducibility and flexibility. & \cite{grant2020kubeflow} \\
\hline
MLOps for Healthcare Diagnostics & Employs microservices for image preprocessing, feature extraction, and AI-based diagnosis in distributed hospital networks, improving data isolation and privacy compliance. & \cite{testi2024machine} \\
\hline
Financial Fraud Detection Systems & Uses microservices for data stream ingestion, feature engineering, and ensemble model serving. Real-time scoring and alerting are achieved through asynchronous event-driven architecture. & \cite{teodoras2024implementing} \\
\hline
Autonomous Vehicles and Sensor Fusion & Adopts microservices for real-time sensor data fusion, path prediction, and decision-making modules. Each sensor stream is processed by independent containerized services to reduce latency. & \cite{ramos2024enhancing}\\
\hline
Retail Demand Forecasting Systems & Integrates microservices for data ingestion, feature processing, and model retraining to handle non-stationary demand patterns. Improves update frequency and model monitoring. & \cite{zacharias2025event} \\
\hline
Hybrid Cloud ML Systems & Applies microservices to distribute training and inference workloads across cloud and edge nodes, improving data locality and reducing latency for hybrid environments. & \cite{rasaq2025resource} \\
\hline
Event-Driven ML in IoT Platforms & Utilizes microservices to trigger inference pipelines based on IoT event streams. Edge microservices perform lightweight inference before syncing with cloud services for retraining. & \cite{ponnaganti2025scalable} \\
\hline
Serverless ML Deployment (FaaS) & Decomposes ML tasks into small function-based services for inference. Reduces operational overhead and enables auto-scaling in cloud-native ML systems. & \cite{loconte2024serverless} \\
\hline
Telecom Network Optimization & Uses distributed microservices to collect, analyze, and predict network traffic patterns using ML algorithms. Facilitates on-demand model updates and resource allocation. & \cite{vemulatransforming} \\

\hline
\end{tabular}
\label{tab:microservices_review}
\end{table*}

Netflix uses a microservice-based architecture to run its large-scale recommendation and personalization systems~\cite{amatriain2015recommender}. Instead of building one large application, Netflix divides its recommendation pipeline into smaller, independent services, each responsible for a specific task such as collecting data, running model predictions, or ranking results. These services communicate through APIs, and an API gateway manages how requests are routed, authenticated, and distributed across the system~\cite{amatriain2015recommender}. To handle the massive flow of user interactions happening in real time, Netflix uses event-driven tools like Kafka and RabbitMQ, which allow services to react dynamically as new data arrives~\cite{mayank2025event}. Reliability and communication between services are maintained using tools like Istio and Linkerd, which manage traffic, balance workloads, and recover quickly from failures~\cite{oye2024microservices}. Together, these technologies show how a microservice approach helps Netflix build scalable, resilient, and flexible machine learning systems. This modular design makes it easier to test, update, and innovate continuously while keeping performance high. Table \ref{tab:microservices_review} provides an overview of major research and industrial work that applies similar principles across different machine learning domains. 
\section{System Design Problem}
Many organizations find it difficult to manage and grow their machine learning systems as they become more complex. When everything is built together in one large application, changing or fixing a single part can affect the whole system. This makes updates slow and expensive. It also becomes difficult to test, reuse, or scale parts that need more computing power. By dividing the system into smaller pieces that work on their own, teams can make changes faster and run updates without breaking other parts. This paper discusses how using microservices and containerization can make machine learning systems easier to build, maintain, and scale in a reliable way.

\begin{figure*}[!t]
    \centering
    \includegraphics[width=\linewidth]{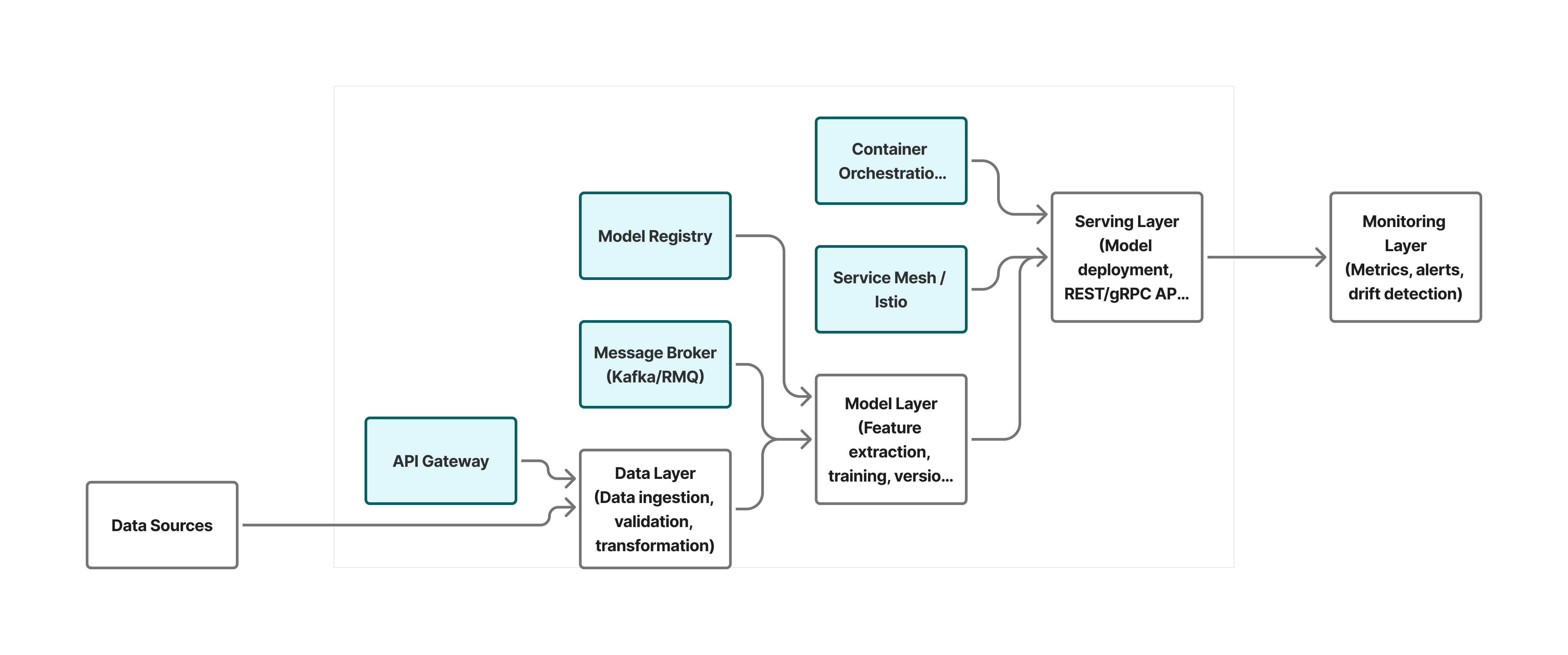}
    \caption{Components of a modular microservice-based machine learning application}
    \label{fig:proposed_arch}
\end{figure*}

\textit{Challenges:}
Building machine learning systems with microservices can make them more flexible, but it also brings some real-world problems. As the number of services increases, it becomes harder to keep everything working together smoothly. When something goes wrong, it can be difficult to find which part caused the issue. Each service may use different tools or versions, so they need careful coordination to avoid conflicts. Security is another big concern, as every service must handle data safely and prevent unauthorized access. Running a lot services can also use more resources and increase costs if they are not managed well.

\textit{Anti-Patterns:}
There are some common mistakes that make microservice-based machine learning systems harder to maintain. One issue is writing quick scripts to connect different parts of the system instead of creating clear and reusable links. Ignoring version control for models or data can lead to errors and make it hard to track changes. Poor communication between services can result in missing or mismatch data, which affects the quality of model predictions. Another mistake is putting too many functions into one service, which reduces the flexibility that microservices are meant to provide. Avoiding these problems helps keep systems reliable, simple, and easier to improve.

\begin{tcolorbox}[
  colback=blue!5,
  colframe=blue!70!black,
  title=\textbf{System Design Problem}
]
\textit{Can we design microservice architectures for machine learning applications that scale gracefully while maintaining reliable service coordination, consistent model and data versioning, secure communication, and efficient resource utilization, without introducing architectural anti-patterns that compromise system maintainability and prediction quality?}
\end{tcolorbox}

\section{Modular Microservice Framework for Machine Learning Systems}

The proposed solution introduces a modular microservice framework (see Figure \ref{fig:proposed_arch}) that redefines how machine learning systems are designed, deployed, and managed in production. Instead of relying on a single monolithic application that handles all parts of the machine learning workflow, the system is divided into a collection of smaller, self-contained services. Each service is responsible for a specific stage of the lifecycle such as data ingestion, preprocessing, feature extraction, model training, inference, or performance monitoring. These services communicate through well-defined APIs, ensuring that each component can evolve independently while remaining part of a cohesive pipeline. This modularity allows teams to upgrade or replace individual services without affecting the stability of the entire system, promoting agility, scalability, and fault isolation.

At the core of this framework is a layered structure that follows natural flow of a machine learning system. The data layer handles collecting, cleaning, and transforming raw data so it’s ready for analysis. This layer uses containerized ETL pipelines that can automatically scale depending on how much data is coming in or how complex the processing is. Next, the model layer focuses on feature engineering, model training, and version control. Each training job runs inside its own container to make sure the environment is consistent and the results are easy to reproduce. Secure APIs allow these services to access shared datasets safely and efficiently. Once a model has been trained and validated, it moves to the serving layer, where it’s deployed as a lightweight, independent service that can be reached through REST or gRPC APIs. Kubernetes manages this layer by handling scaling, load balancing, and failover, ensuring that the system stays responsive and reliable even under heavy workloads. Finally, the monitoring layer monitorsn how the system is performing in real time. It collects metrics, detects anomalies, and checks for issues like data drift. Tools like Prometheus and Grafana make it easy for developers and operators to visualize performance, identify problems quickly, and keep the entire system running smoothly.

This framework also integrates several proven design concepts adapted from software engineering to address challenges unique to machine learning. An API gateway serves as the single entry point for all client interactions, which simplifies authentication and routing. Event-driven communication using message queues such as Kafka or RabbitMQ ensures asynchronous coordination between components, reducing dependencies and improving resilience. A centralized model registry maintains metadata, training configurations, and performance scores, which allow for traceability and safe rollbacks during experimentation. The sidecar pattern is applied to offload logging, monitoring, and security functions to auxiliary containers, keeping the core service logic clean and maintainable. Additionally, a service mesh like Istio provides advanced traffic management, encryption, and inter-service observability, which further improving reliability and fault tolerance.

The real strength of this approach is its flexibility and ease of maintenance over time. Because the system is built from separate, independent components, teams can update models or adjust data pipelines without interrupting the live production environment. When something goes wrong, the issue is contained within a single service, making it much easier to fix and recover quickly. Containerization also helps by keeping every stage of development—whether in research, testing, or production—consistent and predictable, reducing the usual headaches that come with deployment. Another major benefit is cost efficiency: since each service can scale up or down based on its own workload, resources are used more effectively, avoiding the waste and rigidity often seen in large, monolithic systems.

In summary, the proposed modular microservice framework offers a practical and flexible way to build and manage modern machine learning systems. It brings structure and clarity to complex workflows while still allowing room for continuous growth and adaptation. By combining solid engineering principles with the fast-changing nature of data-driven applications, this approach helps create ML systems that are reliable, easy to reproduce, and capable of evolving to meet the needs of large-scale, real-world deployments.

\section{Case Study - Netflix Recommendation System}
In this section, we present a detailed overview of the Netflix recommendation system architecture \cite{amatriain2015recommender} and examine how microservices play a key role in its design and scalability. The Netflix platform serves millions of users worldwide and depends heavily on machine learning to personalize content recommendations in real time. To manage this level of complexity and traffic, Netflix has adopted a distributed architecture where the entire recommendation workflow from data processing and model training to serving and monitoring is divided into smaller, independent components. We discuss how these components operate across the offline, nearline, and online layers of the system and explain how microservices enable flexibility, faster updates, and seamless scaling across each layer.

The Netflix recommendation system is built around three main layers: the offline, nearline, and online components. Each layer plays a specific role in how recommendations are created, refined, and served to users. They work together to form a complete cycle where data flows from large-scale analysis to real-time decision making. Thinking of this setup through the lens of microservices makes it easier to see how each layer can function as an independent unit that can be built, deployed, and scaled without affecting the rest of the system. This approach not only improves flexibility but also helps teams work faster and with fewer system-wide failures.

The offline layer forms the base of Netflix’s recommendation engine. It deals with collecting and processing massive amounts of user data and using that data to train machine learning models. Information about what users watch, search for, and rate is stored in large data systems like Hadoop and Hive. This data is then cleaned, processed, and transformed into a usable format. In a microservice-based setup, each of these steps could be handled by separate services. One service might take care of ingesting and cleaning data, another could extract features or create summary statistics, and a dedicated training service could build and evaluate models. Once a model is ready, it can be stored and versioned by a model registry service so that future deployments are smooth and traceable. Containerization makes these services portable and reliable, allowing teams to train or retrain models independently without disrupting the entire pipeline.

The nearline layer acts as a bridge between the large batch computations done offline and the real-time personalization that happens online. It processes user actions like playing a show, adding to a list, or giving a rating soon after they occur. These updates help keep recommendations fresh without retraining the entire model from scratch. In a microservice setup, this could be handled by event-driven services that respond to user actions as they happen. A streaming service could collect user activity, while another service could update scores or recalculate short-term preferences. Caching services could store recent results, and a publishing service could send updated recommendations to the online system. These services run independently and communicate through event streams, which allows the system to stay responsive and resilient even when user activity spikes.

\begin{figure}
    \centering
    \includegraphics[width=0.9\linewidth]{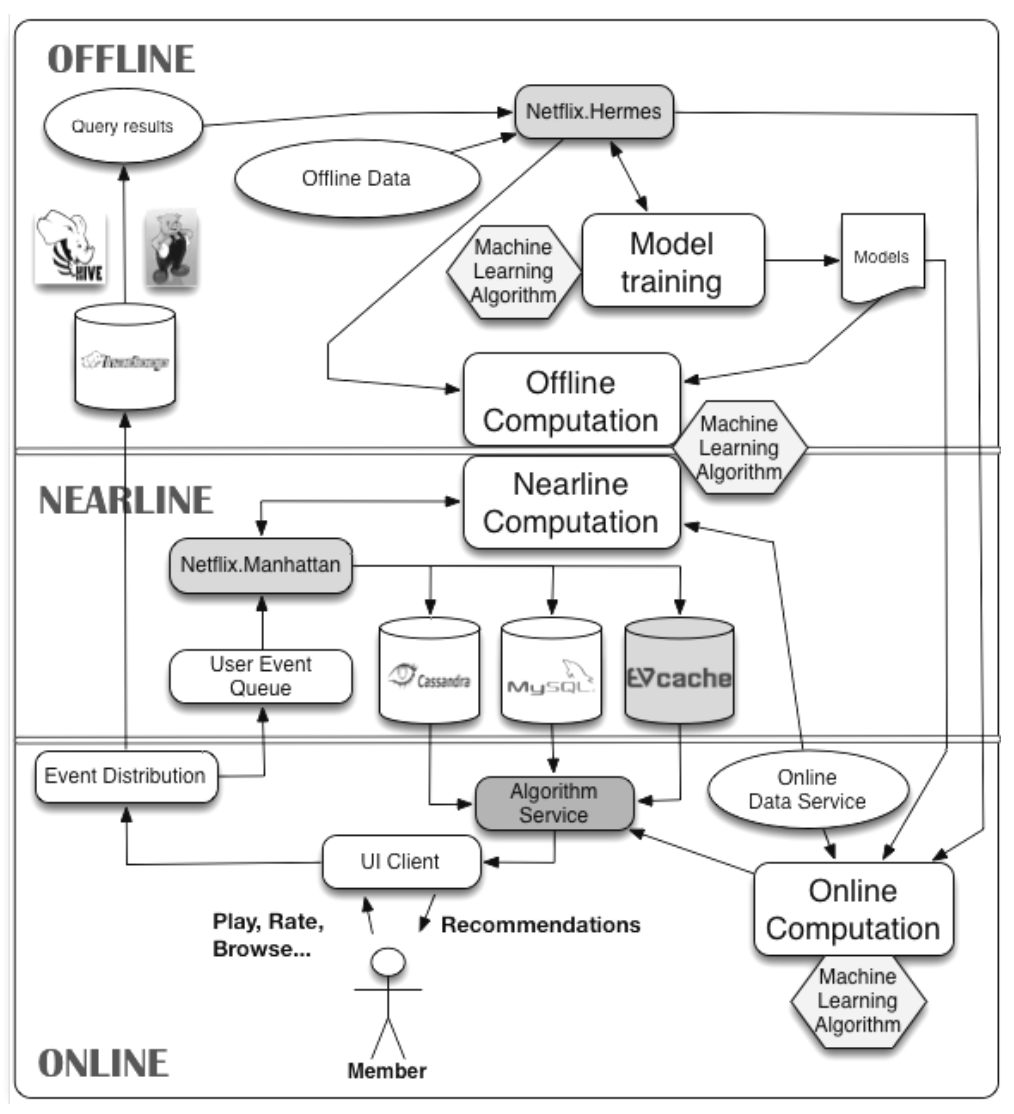}
    \caption{Architecture Diagram of Netflix Recommendation System \cite{amatriain2015recommender}}
    \label{fig:placeholder}
\end{figure}
The online layer is the part users interact with directly when browsing or watching content. It delivers personalized recommendations almost instantly based on the latest available data. This layer is made up of fast, lightweight services that handle inference, ranking, and presentation. For example, an inference service could generate a list of potential shows or movies, while a ranking service determines the best order for each user. An API service connects these outputs to the Netflix app or web interface. Because these services are independent and stateless, they can easily scale up during high-traffic times and scale down when usage is low. Monitoring services also track system health and user behavior, feeding this information back into the nearline and offline layers to improve future recommendations.

By structuring the system around microservices, Netflix can manage each stage of its recommendation process separately while keeping the entire system consistent and efficient. This modular approach helps the company roll out new models or algorithms faster, recover quickly from failures, and maintain reliable service for millions of users. It creates a living system that constantly learns and improves while staying flexible enough to adapt to new technologies and user needs.

\section{Containers - The Backbone for Microservice Deployments}

Containerization has completely changed the way machine learning systems are built and deployed. It serves as the backbone for microservice-based architectures, making it possible for different parts of a machine learning workflow to run smoothly together without depending on a single system setup~\cite{zhong2022machine}. Each task in the pipeline, whether it is data cleaning, feature generation, training, or prediction, can run inside its own small, self-contained environment. These environments, called containers, carry everything they need to work properly, including code, libraries, and configuration files. This means that once something works in one place, it will work the same way everywhere else, from a developer’s laptop to a large production cluster.

Technologies like Docker and Kubernetes have made this approach both reliable and practical. Docker makes it easy to package each part of an application into a portable container, while Kubernetes helps manage how these containers run across many computers. Together, they make it simple to launch, scale, and update systems that might involve hundreds or even thousands of services. With containerization, machine learning teams can test new models quickly, fix issues without bringing the whole system down, and make changes without worrying about breaking other components. It has turned complex ML operations into something far more manageable and predictable.

Another big advantage of containerization is how it supports high-performance computing for machine learning. Many modern workloads rely on GPUs to train large models, and specialized container systems now make it easy to run these heavy computations in parallel across multiple machines. These “supercomputing containers” let engineers train deep learning models faster while keeping the environment consistent and portable. They combine the flexibility of cloud-native tools with the raw power of GPU clusters, bridging the gap between research experiments and real-world deployment.

Today, containerization is used almost everywhere, from tech giants running global recommendation systems to startups deploying custom AI solutions. It has become the common ground that connects software engineering with large-scale machine learning, allowing teams to move faster, scale confidently, and deliver reliable, production-ready AI and ML systems.

\section{Simulations and Results}

In this section, we study the effectiveness of microservice architecture in improving the scalability and responsiveness of machine learning systems. The goal of these simulation studies is to mathematically model and measure how microservices impact key system performance metrics such as latency and throughput. By comparing these results with traditional monolithic setups, we aim to highlight the quantifiable advantages that microservices bring to large-scale, real-time ML applications.

We consider a simplified recommendation system to illustrate the scalability implications of architectural design choices. The system consists of $n$ users, where each user $i \in \{1,\dots,n\}$ is associated with a preference vector $u_i \in \mathbb{R}^d$. Given an item feature vector $v \in \mathbb{R}^d$, recommendations are generated using a standard inner-product scoring function $s_i = u_i^\top v$. Since the focus of this experiment is on system-level behavior rather than model complexity, the computational cost of this scoring operation is treated as constant and identical across all architectures. We compare two deployment scenarios: a \emph{monolithic architecture}, in which all user preference vectors are stored in a centralized service and accessed over the network for every recommendation request, and a \emph{microservice architecture}, in which each user (or a small shard of users) maintains a locally stored preference vector co-located with the recommendation logic. The primary objective is to examine how end-to-end recommendation latency evolves as the number of users in the system increases.
\begin{figure}[!h]
    \centering
\includegraphics[width=0.9\linewidth]{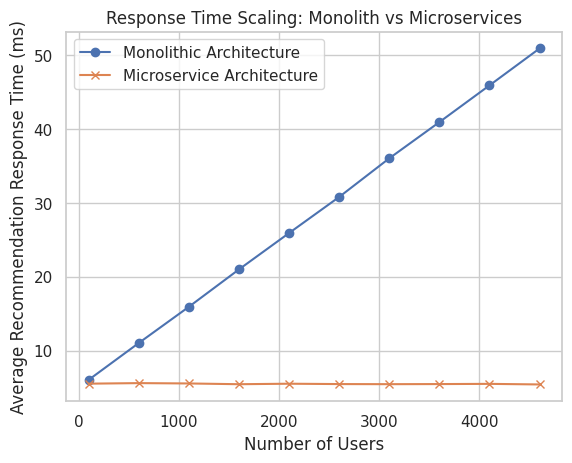}
    \caption{Response Time Scaling: Monolith vs Microservices for Recommendation Systems}
    \label{fig:exp}
\end{figure}

\textit{Simulation Methodology: }
To evaluate scalability, we vary the number of users $n$ over a predefined range and simulate recommendation requests under both architectures. For each value of $n$, we perform multiple independent trials and report the average response time. In the monolithic setting, response time is modeled as
\begin{equation}
T_{\text{mono}}(n) = T_{\text{comp}} + \alpha n + \varepsilon,
\end{equation}
where 
\begin{itemize}
    \item $T_{\text{comp}}$ -- is a fixed base computation time,
    \item  $\alpha n$  -- captures network latency and contention effects that grow linearly with the number of users, and
    \item $\varepsilon \sim \mathcal{N}(0,\sigma_{\text{mono}}^2)$ -- is Gaussian noise modeling stochastic system variability.
\end{itemize}
   In the microservice setting, response time is modeled as
\begin{equation}
T_{\text{micro}} = T_{\text{comp}} + T_{\text{local}} + \varepsilon,
\end{equation}
where
\begin{itemize}
    \item $T_{\text{comp}}$ -- is a fixed base computation time,
    \item  $T_{\text{local}}$  -- captures local computation latency, and
    \item $\varepsilon \sim \mathcal{N}(0,\sigma_{\text{micro}}^2)$ -- is Gaussian noise modeling stochastic system variability.
\end{itemize}

We have $\sigma_{\text{micro}} < \sigma_{\text{mono}}$, reflecting reduced variability due to local access and the absence of centralized contention. For each configuration, response times are averaged across trials to smooth random fluctuations.

\textit{Results and Interpretation: }
The results from our experiments show a clear difference between the two system designs (see Figure \ref{fig:exp}). As the number of users increases, the monolithic setup slows down steadily because every request goes through the same central process and shared data sources. In contrast, the microservice-based system keeps its response time almost the same, even as the workload grows. This happens because each service handles its own part of the work and stores the information it needs locally, reducing unnecessary communication and avoiding bottlenecks. Although our experiment uses a simplified model, the pattern we observed is consistent with what happens in large-scale systems used in practice. These results emphasize that the overall system design, not just the machine learning model itself can have a major impact on how well a system scales and how fast it responds to users.

\section{Conclusion}
In this paper, we looked at how microservice architectures are changing the way machine learning systems are built and managed. We started by discussing the main ideas behind microservices and containerization, focusing on how they make complex ML workflows easier to scale, maintain, and improve over time. Our review included examples from real-world systems where companies have adopted these approaches to boost performance and flexibility. We also took a closer look at the Netflix recommendation system to see how microservices fit into a large-scale ML environment, helping it deliver real-time recommendations, faster updates, and smoother coordination between different components. To test these ideas further, we carried out simulation studies comparing microservice-based and monolithic architectures. The results showed that microservices offer clear advantages in scalability, speed, and overall system efficiency. Altogether, our findings show that microservices are not just a design choice—they are a practical and effective way to build machine learning systems that can grow, adapt, and perform reliably at scale.

\bibliographystyle{IEEEtran}
\bibliography{ref.bib}

@article{hermann2018scaling,
  title={Scaling machine learning at uber with michelangelo},
  author={Hermann, Jeremy and Del Balso, Mike and Kjelstr{\o}m, K{\aa}re and Reinhold, Emily and Beinstein, Andrew and Sumers, Ted},
  journal={Uber Engineering, Nov},
  volume={2},
  year={2018}
}

@incollection{amatriain2015recommender,
  title={Recommender systems in industry: A netflix case study},
  author={Amatriain, Xavier and Basilico, Justin},
  booktitle={Recommender systems handbook},
  pages={385--419},
  year={2015},
  publisher={Springer}
}

@inproceedings{baylor2017tfx,
  title={Tfx: A tensorflow-based production-scale machine learning platform},
  author={Baylor, Denis and Breck, Eric and Cheng, Heng-Tze and Fiedel, Noah and Foo, Chuan Yu and Haque, Zakaria and Haykal, Salem and Ispir, Mustafa and Jain, Vihan and Koc, Levent and others},
  booktitle={Proceedings of the 23rd ACM SIGKDD international conference on knowledge discovery and data mining},
  pages={1387--1395},
  year={2017}
}

@article{mayank2025event,
  title={Event-Driven Architectures for Serverless ETL: Redefining Data Pipeline Reactivity in Cloud-Native Environments},
  author={Mayank, Chauhan and Singh, VK},
  year={2025}
}

@article{zhong2022machine,
  title={Machine learning-based orchestration of containers: A taxonomy and future directions},
  author={Zhong, Zhiheng and Xu, Minxian and Rodriguez, Maria Alejandra and Xu, Chengzhong and Buyya, Rajkumar},
  journal={ACM Computing Surveys (CSUR)},
  volume={54},
  number={10s},
  pages={1--35},
  year={2022},
  publisher={ACM New York, NY}
}

@article{oye2024microservices,
  title={Microservices Architecture for Large-Scale AI Applications},
  author={Oye, Emma and Frank, Edwin and Owen, Jane},
  year={2024}
}

@article{palamarchuk2022methods,
  title={Methods of building microservice architecture of e-learning systems},
  author={Palamarchuk, Ye A},
  year={2022},
  publisher={ВНТУ}
}

@article{shabani2022design,
  title={Design of modern distributed systems based on microservices architecture},
  author={Shabani, Isak},
  journal={International Journal of Advanced Computer Science and Applications},
  year={2022}
}

@inproceedings{de2019monolithic,
  title={From monolithic architecture to microservices architecture},
  author={De Lauretis, Lorenzo},
  booktitle={2019 IEEE International Symposium on Software Reliability Engineering Workshops (ISSREW)},
  pages={93--96},
  year={2019},
  organization={IEEE}
}

@article{blinowski2022monolithic,
  title={Monolithic vs. microservice architecture: A performance and scalability evaluation},
  author={Blinowski, Grzegorz and Ojdowska, Anna and Przyby{\l}ek, Adam},
  journal={IEEE access},
  volume={10},
  pages={20357--20374},
  year={2022},
  publisher={IEEE}
}

@article{testi2024machine,
  title={Machine Learning Operations (MLOps) in Healthcare},
  author={Testi, Matteo},
  year={2024},
  publisher={Universit{\`a} Campus Bio-Medico}
}

@inproceedings{teodoras2024implementing,
  title={Implementing a Java Microservice for Credit Fraud Detection Using Machine Learning},
  author={Teodoras, Dan-Alexandru and Stalidi, Cosmina and Popovici, Eduard-Cristian and Suciu, George},
  booktitle={2024 23rd RoEduNet Conference: Networking in Education and Research (RoEduNet)},
  pages={1--5},
  year={2024},
  organization={IEEE}
}

@inproceedings{ramos2024enhancing,
  title={Enhancing autonomous vehicles control: Distributed microservices with V2X integration and perception modules},
  author={Ramos, Joaquim and Figueiredo, Andreia and Almeida, Pedro and Aston, Tiago and Campos, Andr{\'e} and Perna, Gon{\c{c}}alo and Mendes, Marcos and Rito, Pedro and Sargento, Susana},
  booktitle={2024 IEEE International Conference on Mobility, Operations, Services and Technologies (MOST)},
  pages={143--153},
  year={2024},
  organization={IEEE}
}

@article{zacharias2025event,
  title={Event-Driven Microservices: Powering Real-Time Retail Innovation},
  author={Zacharias, Juby Nedumthakidiyil},
  journal={Journal of Computer Science and Technology Studies},
  volume={7},
  number={6},
  pages={24--31},
  year={2025}
}

@article{rasaq2025resource,
  title={Resource Optimization in Hybrid Cloud: Leveraging Microservices for IoT Workloads},
  author={Rasaq, SODIQ OYETUNJI},
  journal={International Journal of Novel Research in Engineering \& Pharmaceutical Sciences},
  volume={1},
  number={1},
  year={2025}
}

@inproceedings{ponnaganti2025scalable,
  title={Scalable Multi-Model Orchestration in AI Microservices with Kubernetes and Serverless for Event-Driven MLOps Pipelines},
  author={Ponnaganti, Venkata Thilak},
  booktitle={2025 International Conference on Intelligent Computing and Control Systems (ICICCS)},
  pages={1471--1476},
  year={2025},
  organization={IEEE}
}

@article{loconte2024serverless,
  title={Serverless microservice architecture for cloud-edge intelligence in sensor networks},
  author={Loconte, Davide and Ieva, Saverio and Gramegna, Filippo and Bilenchi, Ivano and Fasciano, Corrado and Pinto, Agnese and Loseto, Giuseppe and Scioscia, Floriano and Ruta, Michele and Di Sciascio, Eugenio},
  journal={IEEE Sensors Journal},
  year={2024},
  publisher={IEEE}
}

@article{vemulatransforming,
  title={TRANSFORMING TELECOM: THE IMPACT OF MICROSERVICES ARCHITECTURE},
  author={Vemula, Krishna Rao}
}

@article{olowoniyianalyzing,
  title={Analyzing Uber’s MLOps Journey: Lessons from QA Failures in Real-Time Recommendation Engines},
  author={Olowoniyi, Rawlings and Kamarapu, Karthik and Hancock, Pearl}
}

@article{singh2024development,
  title={From development to deployment: Streamlining MLOps with monoliths microservices and Amazon SageMaker},
  author={Singh, Karanbir},
  journal={International Research Journal of Engineering and Technology},
  volume={11},
  number={9},
  pages={16},
  year={2024}
}

@book{barnes2015microsoft,
  title={Microsoft Azure essentials Azure machine learning},
  author={Barnes, Jeff},
  year={2015},
  publisher={Microsoft Press}
}

@book{grant2020kubeflow,
  title={Kubeflow for machine learning},
  author={Grant, Trevor and Karau, Holden and Lublinsky, Boris and Liu, Richard and Filonenko, Ilan},
  year={2020},
  publisher={" O'Reilly Media, Inc."}
}


\end{document}